\documentclass{physeauth}
\usepackage{graphicx}
\usepackage{amsmath}
\usepackage{amssymb}

\usepackage{mymacros}

\begin{document} 

\begin{frontmatter}

\title{Quantum Ratchets at High Temperatures}

\author{D. Zueco \thanksref{thank1}} and 
\author{J. L. Garc\'{\i}a-Palacios}

\address{Dep. de F\'{\i}sica de la Materia Condensada e Instituto
de Ciencia de Materiales de Arag\'on.  
\\
C.S.I.C.-Universidad de Zaragoza.  E-50009 Zaragoza. Spain }

\thanks[thank1]{
Corresponding author. 
E-mail: zueco@unizar.es}

\begin{abstract}
 Using the continued-fraction method we solve the {\it Caldeira-Leggett}
master equation in the phase-space ({\it Wigner}) representation to study
Quantum ratchets.  
Broken spatial symmetry, irreversibility and periodic forcing allows
for a net current in these systems.
We calculate this current as a function of the force under adiabatic
conditions.
Starting from the classical limit we make the system
quantal.  
In the quantum regime tunnel events and over-barrier wave
reflection phenomena modify the classical result.
Finally, using the phase-space formalism we give some insights about the
decoherence in these systems.
\end{abstract}

\end{frontmatter}


\section{INTRODUCTION}

Transport properties in periodical structures with broken spatial
symmetry (ratchet systems) have attracted a lot of attention in the past
decade \cite{rei02}.  These systems, under out of equilibrium
conditions, can display net current even when the time and space
average of
all applied  forces is zero.  For this reason, these systems are also
called {\it Brownian} motors
\cite{asthan02}. Realizations of {\it Brownian} motors can be
found, for example,
in biological and condensed matter systems \cite{lined02}.

In the classical limit, the dynamics is governed by classical {\it
Langevin} or equivalently {\it Fokker-Planck} equations.  This limit
has been extensively studied in the recent past, establishing the main
phenomenologies of the ratchet effect (non vanishing rectified
current, etc...)  \cite{rei02}.  The inclusion of quantum fluctuations
in the system enriches the ratchet phenomenology \cite{reigrihan97}.
However, the difficulties to deal with quantum dissipative systems,
make their study harder and one finds less works in this regime \cite
{lined02,reigrihan97,schvin02,yuketal97,grietal02,macetal04,peggri04}. 

Recently, it has been shown that it is possible to solve Quantun
Master Equations using continued fraction methods
\cite{gar04,garzue04}. 
Such Master Equations are the ``quantum
generalization'' of the classical {\it Fokker-Planck} equations for
some ranges of the parameters (weak coupling, high
temperature,...). Adapting this technique from the classical case
\cite{gar04}, the authors of  \cite{garzue04} have
studied the ratchet effect as a function of the temperature.  In this
work we extend the study to the force dependence of the rectified
velocities in the high temperature/weak coupling regime.


\section {THEORETICAL FRAMEWORK}
\label{theoretical}

The out of equilibrium dynamics of classical systems is well
established since one century ago with the seminal works of {\it Einstein}
and {\it Langevin}. 
The {\it Langevin} ({\it
Fokker-Planck}) equations describe satisfactorily the stochastic
behavior of these systems.  However, their quantization represents in
most cases a difficult task \cite{weiss}.  The most satisfactory
approach consists of quantizing the Hamiltonian of the System plus its
Environment.  The minimal Bath model is a  large collection of
harmonic oscillators. This approach, in the classical
limit, recovers the {\it Langevin} description of open systems.

In the quantum realm a closed evolution
equation is not possible in the general case.
%
This is mainly due the non-{\it Markovian} nature of the
quantum correlations. 
%
Nevertheless, under some approximations, it is possible to derive a
{\it Markovian} Quantum Master Equation.  
Thus, it is typically assumed that the
relaxation times are much longer that the quantum correlations of the bath,
\begin{equation}
\label{validity}
\frac {1}{\gamma}
\gg
\frac {\hbar}{T}
,
\end{equation}
with $\gamma$ the damping parameter measuring the strength of the
coupling to the bath.
Under these conditions
(high temperatures and/or weak
coupling) the evolution is given by the celebrated {\it
  Caldeira-Leggett} master equation \cite{calleg83pa},
%
%
\begin{equation}
\label{C-L}
i
\hbar
\partial_t
\varrho
=
[ H_S, \varrho]
+
\frac {\gamma}{2 \hbar} 
[\x, \{ \p, \varrho \}]
-
\frac {i \gamma m \kT} {\hbar}
[\x, [\x, \varrho]]
.
\end{equation}
Here $\varrho$ is the reduced density matrix, obtained after
tracing out the environmental degrees of freedom and $H_s$  is the system
Hamiltonian, in our case $H_S = \p^2/2m + V(\x)$.  The first term in
(\ref{C-L}) gives the unitary evolution, the second 
yields the dissipation and the last is the responsible for the
diffusion.

An alternative description of quantum systems is provided by the
phase-space ({\it Wigner}) formalism.  The central object is the {\it
Wigner} function, defined in terms of the density matrix as
\cite{hiletal84}
%
\begin{equation}
\label{wigner:def}
\Wf(\x,\p)
=
\frac{1}{2\pi\hbar}
\int
\drm y\,
\e^{-\iu \p\,y/\hbar}
\varrho(\x+ y/2 ,\x- y/2)
\;.
\end{equation}
This representation not only gives a phase-space description for
quantum systems, but it allows to extend classical concepts and tools
into the quantum domain. Besides, it permits to find quantum analogues
to classical phenomena (Quantum Chaos, phase-space trajectories, ...)
and provides a natural Quantum-Classical connection.

In the {\it Wigner} representation the master equation (\ref{C-L}), can be written as 
\begin{eqnarray}
\label{WKK:scaled}
\nonumber
\partial_{\tT}
\Wf
=
\Big[
-\p\,
\px
+
\bV'\,
\pp
+
\gammaT\,
\pp\,
\big(\p+\pp\big)
\\  
+
{\textstyle\sum}_{\iq=1}^{\infty}
\qcoef^{(\iq)}
\,
\bV^{(2\iq+1)}(\x)
\,
\pp^{(2\iq+1)}
\Big]
\Wf
\;.
\end{eqnarray}
The first two terms correspond to the {\it Poisson} bracket ({\it
  Liouville} equation). This {\it Poisson} bracket plus the third term
  yields the classical {\it Fokker-Planck} equation, evidencing the
  classical character of the noise and relaxation.  The last term is
  purely quantum and comes from the unitary evolution of the closed
  system ({\it Wigner-Moyal} bracket).

Equation (\ref{WKK:scaled}) is written in dimensionless form
\cite{garzue04}.  The variables have been scaled with help from a
reference length $x_0$, mass $M$ and frequency $\Omega_0$.  For
example, action variables are scaled by the characteristic action $S_0
= M\Omega_0x_0^2$, energy by $E_0 = M\Omega_0^2x_0^2$, etc. Then the
coefficient in the quantum sum is given in terms of the {\it de
Broglie} wave length by,
%
\begin{equation}
\label{kappa:s}
\qcoef^{(\iq)}
=
(-1)^{\iq}\ldB^{2\iq}\big/(2\iq+1)!
\;.
\end{equation}
Besides, $\hbar$ is introduced in terms of $S_0$ via the quantum parameter $K$:
%
\begin{equation}
\label{kondos}
\hbar/\So=2\pi/\kondobar
\;
\quad 
\mbox{and}
\qquad
\ldB \propto 1/\kondobar
\;.
\end{equation}
Note that the classical limit is naturally recovered letting $K
\rightarrow \infty$.

To conclude, the calculation of observables in the {\it Wigner}
formalism is an example of extension of classical methods to the
quantum.  The Quantum Mechanical expectation value is obtained
via $W(\x,\p)$ as a classical ``average''
%
\begin{equation}
\label{average}
\big\langle A\big\rangle
\equiv
\mathrm{Tr}(\hat{\varrho}\,\hat{A})
=
\int\!\drm\x\drm\p\,
\Wf(\x,\p)\,
A(\x,\p)
\;,
\end{equation}
with $A(\x,\p)$ being the classical observable corresponding to the
operator $\hat {A}$ (via Weyl's rule).


\section {CONTINUED FRACTION APPROACH}

A suitable non-perturbative technique to solve classical {\it
  Fokker-Planck} equations of systems with few variables is the
  continued-fraction method \cite{risken}. This is a special case of
  the expansion into complete sets (Grad's) method to solve kinetic
  equations in statistical mechanics.  The technique
  had already been adapted to quantum dissipative systems in studies
  of spins  and quantum nonlinear optics.  Recently this method has been
  extended to quantum Brownian Motion problems described by {\it
  Caldeira-Leggett} type equations exploiting the {\it
  Fokker-Planck}-like structure of the quantum master equation in the
  {\it Wigner} representation \cite{gar04,garzue04}.

The idea of the method consists of expanding the desired
solution $W$ of the differential equation ($\partial_t W = \mathcal
{L} W$) into an appropriate basis.  The equations for the expansion
coefficients ($C_i$) have then the form of a system of coupled differential
equations, say
\begin{equation}
\label{recurrencegeneral}
\dot 
C_j
=
\sum _{i=I}^ {i=-I}
Q _{j, j+i}
\;
C_{j+i}
.
\end{equation}
  The goal is to find a basis in which the range of index coupling $I$
of the coefficients $C_i$ is as short as possible.  Indeed, for 
finite coupling range ($I < \infty$) the differential recurrence
relation (\ref{recurrencegeneral}) can be solved by iterating a simple
algorithm, the structure of which is like that of a continued
fraction,
%
\begin{equation}
\label{continedfraction}
C
=
\cfrac{p_1}{q_1 + 
 \cfrac
{p_2}{q_2 + 
  \cfrac
{p_3}{q_3 +
\dotsb
}}}
\end{equation}

In the case of the {\it Caldeira-Leggett} equation the expansion is on an
$\x$ $\&$ $\p$ basis:
%
\begin{equation}
\label{W:expansion}
\Wf(\x,\p)
\propto
\sum_{\ip, \ix}
\ec_{\ip}^{\ix}
u_{\ix}(\x)
\psi_{\ip}(\p)
\;.
\end{equation}
Then, the recurrence (\ref{recurrencegeneral})  is
replaced by a two index one,
$
\dot C _n ^{\alpha}
=
\sum_{m,\beta} Q_{\alpha \beta} ^{n m}
C_m ^{\beta}
$, 
which can be transformed into a one index  recurrence introducing appropriate
vectors and matrices:
\begin{equation}
\frac {d}{dt} {\bf C} _{\alpha}
=
\sum_{i=-I}^{i=I}
\mathbb {Q} _{\alpha, \alpha + i} {\bf C} _{\alpha +i}
.
\end{equation}
This recurrence relation can be solved with {\em matrix}
continued-fraction methods.  In \cite{garzue04} the explicit form of
matrices $\mathbb {Q}$ was derived.  In particular, for periodic
potentials [$V(\x) = V(\x + L)$] explicit recurrence relations were
constructed using {\it Hermite} functions for the momentum basis and
{\it plane waves} ($ u_{\alpha} \propto \e^{i \alpha x}$) for the
position basis. This choice provides a finite coupling range $I$,
making possible its solution by the abovesketched method. 

Solving the Quantum Master Equation in this way, the {\it Wigner}
function is obtained ({\it i.e.}, the density matrix), so any
observable can be calculated.

\section {RESULTS FOR QUANTUM RATCHETS}

In the context of {\it Brownian} motion the simplest model that
allows for broken spatial symmetry  is the two harmonic
potential, 
%
\begin{equation}
\label{Vrat}
\V(\x)
=
-\Vo
\big[
\sin\x
+
(\rat/2)
\sin(2\x)
\big]
\;,
\end{equation}
(another ratchet systems can be modelised using tight binding models,
rotational spin systems, optical lattices, etc...).  
Potential (\ref{Vrat}) has a steeper side to the left (see
fig. \ref{bands}), which we will call the ``hard''  direction, while we
will refer to the other direction as the ``easy'' one (to the right).

Using the continued-fraction technique we can solve the {\it
Caldeira-Leggett} equation.  In particular for the case of the two
harmonic potential (\ref{Vrat}) the coupling range $I$ is 2 ({\it
  i.e.}, equal to the number of harmonics of the potential).

Our purpose is to study the transport properties in this system. We 
limit ourselves to the case of a square-wave force switching
alternatively between $\pm F$. Then the rectified velocity is defined as,
\begin{equation}
\label{rectified}
\gamma
\Ppro_{\rm r}
= 
\gamma\Ppro_{+F}
+
\gamma\Ppro_{-F}
\;.
\end{equation}
As we  consider adiabatic conditions ($\omega\to0$), the $\Ppro_{\pm F}$
are the corresponding stationary velocities; they can be calculated
immediately with the general result (\ref{average}).

In ref. \cite{garzue04} we considered the temperature dependence of
the ratchet current.
In this article, we study the force dependence of 
$\gamma
\Ppro_{\rm r}$
at fixed temperature.  
A
high enough temperature is chosen to circumvent validity problems of the
master equation (see below).  

\subsection {Classical Limit}
First, it is  convenient to understand the classical phenomenology.
\begin{figure}[!tbh]
\centerline{\resizebox{8.5cm}{!}{%
\includegraphics[angle =-90]{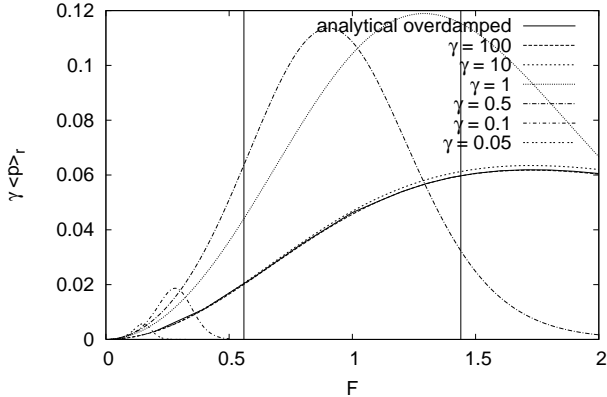}
}}
\caption{
Rectified velocity vs.\ force for a classical particle in a ratchet
potential (a square wave $\pm F$ is used). 
Results are shown for various values of the damping $\gamma$, as well
as the analytical overdamped result.
The vertical lines mark the two critical forces in the deterministic
case $F_c^\pm$ [Eq. (\ref{criticalforces})].
}%
\label{clas}
\end{figure}
We plot in figure \ref{clas} the effect of  finite inertia on the curves $\gamma
\Ppro_{\rm r}$ vs. $F$ at a fixed temperature $\kT/E_0=1$. 


The results obtained can be summarized as follows:
(i) the rectified velocity is positive, {\it i.e.}, the net drift is
to the ``easy'' side;
(ii)
at high $\gamma$ the results converge to the known overdamped
analytical result;
(iii)
finite inertia ($\gamma$) shifts the curves to lower forces and
narrows them;
(iv)
at low enough damping the net current drops to zero.

There are two main points to be understood here: the dependence on $\gamma$
of the efficiency of the rectification and the window of forces where
it occurs.  Recall that in the zero noise limit ($T=0$), there exist
two critical forces (for $r > 1/4$),
\begin{equation}
\label{criticalforces}
|F_c^-/V_0| = 1+ r
,
\qquad
|F_c^+/V_0| = 1- r
,
\end{equation}
where $F_c^\pm/V_0$ is the critical force for barrier dissapperance to
the right and left sides.  In our case $r=0.44$, then $F_c^-/V_0>
F_c^+/V_0$. Thus in the $T=0$ limit we expect a current to the right
({\it i.e.}  positive).

Borromeo and {\it co-workers} studied the classical inertial, deterministic
 ($T=0$) case \cite{borcosmar02}. They found that the net current starts
at $F_c^+/V_0$.  Then 
rectification grows quickly with $F$, up to $|F_c^-/V_0|$. From this force the
rectification starts to
decrease with $F$, 
tending to zero at large forces.
%
They also observed that inertia narrows the curves,
and fits them to the window $(F_c^+/V_0, F_c^-/V_0)$.  
In addition, for 
intermediate damping the ratchet efficiency grows
inversely proportional to the damping. Finally, at very low damping
the rectified velocity drops to zero.  
Thus, our findings agree with the efficiency and narrowing
characteristics of Ref. \cite{borcosmar02} for the noiseless limit.
Nevertheless the temperature smooths the curves in our case, allowing
for rectified velocity below $F_c^+$.

This range below  $F_c^+$ is indeed quite rich already in the
deterministic case.
There are two more characteristic forces $F_1(\gamma)$ and
$F_2(\gamma)$ ($F_1 < F_2 < F_c$).  Below $F_1$ the locked solution
($\langle p \rangle \simeq 0$) is the attractor globally stable.  Above
$F_2$ the stable attractor is the running solution ($\langle p \rangle
\simeq F/\gamma$).  
Between $F_1$ and $F_2$ there exist bistability between these
solutions.  However, both for locked and running solutions there is
little assymetry in the response [as then $\langle p \rangle $ is
nearly independent of $V(x)$].  Therefore, the assymetry in the
response concentrates in the range $F_1 < F< F_2$.
Then, since both $F_1$ and $F_2$ decrease with $\gamma$ \cite[p.~330]{risken},
the window with the maximuun rectification is shifted to lower $F$ as
$\gamma$ is decreased, in accordance with our results.

%


\subsection {Quantum corrections}

\begin{figure}[!tbh]
\centerline{\resizebox{8.5cm}{!}{%
\includegraphics[angle =-90]{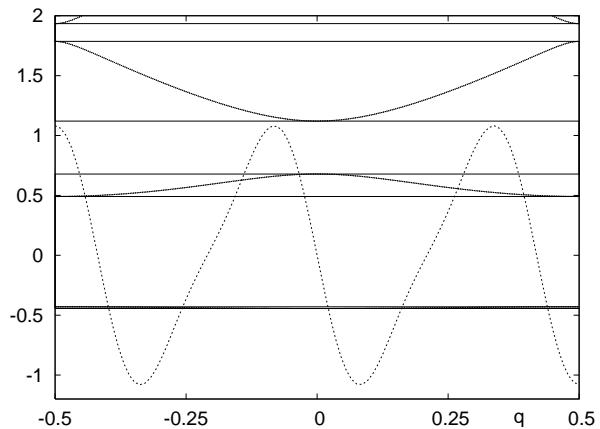}
}}
\caption{Energy bands for $\kondobar=5$.  The potential profile
$
\V(\x)
=
-\Vo
\big[
\sin\x
+
(\rat/2)
\sin(2\x)
\big]
$
 is plotted (dashes) to show the number of bands below the barrier.  In the
  calculations we use $r=0.44$ which smooths the potential profile.
}
\label{bands}
\end{figure}

After the exploration of the classical limit, we proceed to make the
system quantal, by decreasing $\kondobar$ (see
section \ref{theoretical}), and see how this affects the ratchet effect.
The physical meaning of $\kondobar$ can be seen calculating the bands
of the closed system, which depend only on $\kondobar$
\cite{garzue04}. The number of bands grows with $\kondobar$ (as a rule
of thumb the number bands below the potential is approximately
$\kondobar/2$) while the band width increases with decreasing
$\kondobar$. We plot in Fig. \ref{bands} the bands for the case
$\kondobar = 5$.

\begin{figure}[!tbh]
\centerline{\resizebox{8.5cm}{!}{%
\includegraphics[angle =0]{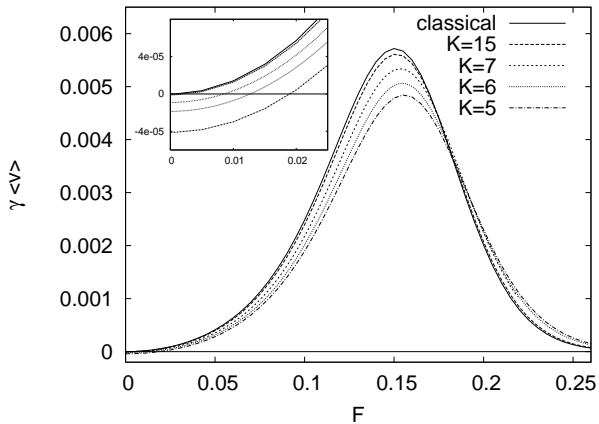}
}}
\caption{
Rectified velocity vs.\ force for a quantum particle in a ratchet
potential (excited with $\pm F$).  
Results are shown for various
values of the quantumness parameter $\kondobar$ (see main text). 
The damping and temperature are fixed at $\gamma = 0.05$ and $\kT/E_0=1$. 
We show for reference the classical case (solid).
Inset: Amplification of the very low force range.  A small negative
rectified current it is observed at $F=0$ for the quantum curves.
}
\label{T1}
\end{figure}

We show results for the rectified current $\gamma \Ppro_{\rm r}$
in Fig. \ref{T1}.  We present curves at $\kT/E_0=1$ and $\gamma =
0.05$ for different values of $\kondobar$. It is observed that the
deviation from the classical case is systematic in $\kondobar$. First,
we find a reduction of the ratchet effect at low forces.  The
reduction continues at intermediate forces, while at high forces an
enhancement is observed, with respect to the classical case.  Finally,
a slight shift of the peaks of the curves to higher forces is
found as $\kondobar$ is decreased.

The reduction/enhancement of the rectified velocity with respect to
the classical limit was already investigated in \cite{garzue04}.
These deviations can be understood in
terms of quasiclassical corrections to  
classical transport, which is
modified by tunnel events and over-barrier wave
reflection (reflection of particles with energy above the barrier;
those energies are attained by thermal activation).
Modelling with an asymmetric {\it saw tooth} potential one finds that
tunnel events are more frequent in the hard  direction that
in the easy one for moderate to weak amplitude forces (leading to reduction of
the ratchet effect)\cite{rei02,garzue04}.  On the contrary,
the phenomenon of the wave reflection is less intense in the easy direction
\cite{garzue04}.

In particular, at $F \simeq \gamma$ we obtained in \cite{garzue04}
that, depending on the inertia and temperature, the reduction/enhancement is a
consequence of the ``competition'' of tunnel events and wave
reflection.  
Increasing the force both the contributions of wave reflection
and tunnel transmission grow \cite
[App.~F]{garzue04}. However wave reflection
grows like $\sim F$, while tunnel transmission goes exponentially with
$F$.  Thus, when $F>\gamma$, but not too high, tunnel
contributions are favored more than wave reflection.  As a
consequence, one finds a reduction of the rectified velocity in the range of
inertia where an enhancement at $F \lesssim \gamma$ takes place.
Finally, at high enough forces the barrier is sufficiently lowered.
Then, there are many particles with energies above the
barrier (experiencing wave reflection).
In addition, at high forces  tunnel events also favor
the enhancement \cite{rei02}.

As for the shift to higher forces, Shushin and Pollak showed 
\cite{shupol03} that quantum corrections produce a suppression of the
diffusion in the underdamped regime.
This effect yields an effective increased barrier, which manifest
itself in a shift to higher forces in the quantum curves.

In figure \ref{T1} (see the inset) we also observe a small current at
$F=0$. We believe that this ``little'' violation of the second law
(transformation of thermal fluctuations in net current in absence of
forcing) is a consequence of the approximations done in the derivation
of the {\it Caldeira-Leggett} equation \cite{macetal04} (since the
starting formalism is thermodynamically consistent).

%


\section {PHASE SPACE REPRESENTATION}

\begin{figure}[!tbh]
\centerline{\resizebox{9.cm}{!}{%
\includegraphics[angle =-90]{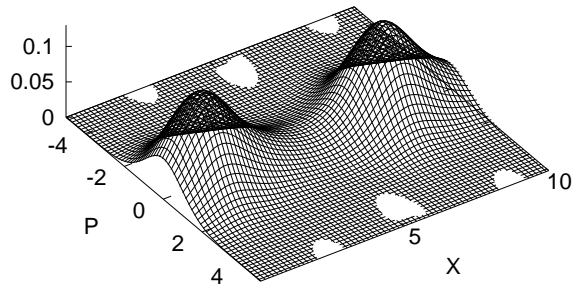}
}}
\caption{{\it Wigner} function for a particle in a ratchet
  potential.  
The quantumness parameter is $\kondobar = 5$, the
  damping is $\gamma = 0.05$  and $F= 0$. 
%
%
White ``islands'' correspond to zones of negative $W$ ($\sim
-10^{-5}$).
%
These negative
zones are of size comparable to those of Hamiltonian case
\cite{garzue04}.
Note the spatial asymmetry  induced by the ratchet potential.
}%
\label{phase-space}
\end{figure}

We plot a  (stationary) {\it Wigner} distribution in fig.\ref
{phase-space}.  In order to see clearly quantum effects we choose the
lowest $K$ case consider ($K \propto S_0 / \hbar$).  
%
%
%

The white islands in the graph represent zones with negative values of
the {\it Wigner} function.  Apart from this negative zones, the {\it
Wigner} distribution lies mainly on its classical counterpart.


Negative zones are a consequence of  quantum  interference. 
Thus, one would expect that decoherence yields positive {\it Wigner}
functions.
In fact, in the harmonic dissipative case, it was shown that after a finite
time the {\it Wigner} function becomes positive
\cite{broozo04}.
However, our plots show that more general potentials allow for non
positive {\it Wigner} distributions at long times.

By decoherence we mean that an intial reduced density matrix,
$\varrho(t_0)$ evolves with time to a diagonal one (in a preferred
basis), when the system is coupled to the environment \cite{zur03RMP}.
The basis $\{|n\rangle \}$ in which $\varrho$ becomes diagonal,
$\varrho = \sum |c_n|^2 |n\rangle \langle n|$, it is the so-called
{\it pointer} basis \cite{zur03RMP}.
The preferred basis depends on the {\it Hamiltonian} of the system
($H_S$) and the interaction between the system and the bath.
For example, in linear Quantum {\it Brownian} Motion (free particle and
harmonic oscillator), the pointer states are localized in phase-space,
the so-called coherent states (familiar from quantum optics).
Mixtures of these states yield positivite $W$.

The fact that our stationary $W$ retains the negative islands (after
decoherence), implies that such coherent states cannot be the pointer
basis in our case.
Actually, in the limit of weak damping and high temperature, one
expects that the stationary solution of (\ref{C-L}) corresponds to the
canonical density matrix,
\begin{equation}
\label{canonical}
\varrho 
=
\frac {1}{\mathcal Z}
\e^{-\beta H_S}
=
\frac {1}{\mathcal Z}
\sum
\e^{-\beta E_k}
|k\rangle \langle k|
\; .
\end{equation}
Accordingly, the pointer states would correspond to {\em eigenstates}
of the system Hamiltonian.

This can be additionally supported by the following.
Consider the approximate Bloch functions that can be analytically
obtained in the extreme quantum case ($K \lesssim 1$) \cite{garzue04}.
Weighting the corresponding $|k\rangle \langle k|$ by the thermal
factor $\e^{-\beta E_k}$ and integrating over $k$'s in the lowest
bands, one obtains Wigner functions with all the structure of the
stationary solutions obtained solving the Caldeira--Leggett equation
for small $\gamma/T$.


\section {CONCLUSIONS}

The continued-fraction method is an established technique to solve
classical {\it Fokker-Planck} equations. Admittedly, compared with
direct simulations, the method has several shortcomings (it is quite
specific of the concrete problem; the stability and convergence fails
in some ranges of parameters, etc...).  However, it also has 
valuable advantages (it is free from statistical errors, its nonperturbative
character, high efficiency, etc...) \cite{risken}.

When transferred to the quantum case the method inherits these
 shortcomings plus the problem of the critical election of the basis
 and recurrence index \cite{garzue04}. However, no quantum {\it
 Langevin} simulations are available, while numerical solutions of
 master equations are computationally  expensive. Notable
 exceptions are the different versions of the quantum stochastic
 calculus \cite{sto03}, which can be viewed as a sort of quantum countenpart
 of the classical {\it Langevin} simulations.  

Let us emphasize that the continued-fraction technique gives the exact
solution of the quantum master equation ({\it Wigner} distribution),
allowing the calculation of any observable [vd.  Eq. (\ref{average})].
This method has also ``spectral advantages'' since it does not require
the calculation of the eigenvalues, circumventing the demanding problem
of continuum spectrum. Besides Quantum-Classical connection is attained in a
natural way (tuned by $\kondobar$). Finally the method  enjoys the
advantages of working in phase-space.

In this work, we have studied the ratchet effect in the quantum case
for high temperature/weak damping conditions. Starting from the
classical case, we have proceed to make the system quantal.  The
deviations from the classical limit have been understood in terms of
tunnel events and thermally activated over-barrier wave reflection.

The representation of the {\it Wigner} functions has shown that, contrary
the harmonic case, ratchet (cosine) potentials do not exhibit positive {\it
Wigner} functions at long times. 
We have argued that in the weak damping and high temperature regime
the pointer basis are the {\it eigenstates} of the system {\it
  Hamiltonian}, yielding in our case a non positive $W$.


We finally mention that the limitations of the master equation
considered do not allow us to study the low temperature regime, where
quantum corrections will be more important. These problems are brought
to the fore with an unphysical violation of the second law of
Thermodynamics.  This situation constitutes a motivation to develop
corrections to the {\it Caldeira-Leggett} equation, which could
hopefully be solved with the continued fraction method.


\section*{Acknowledgments}

This work was supported by DGES, project BFM2002-00113, and DGA, grant
no.\ B059/2003.
We acknowledge to P. J. Mart\'{\i}nez his support during this work.


\bibliography{/home/david/notas_articulos/cosas_tex/david}

\begin{thebibliography}{10}
\expandafter\ifx\csname url\endcsname\relax
  \def\url#1{\texttt{#1}}\fi
\expandafter\ifx\csname urlprefix\endcsname\relax\def\urlprefix{URL }\fi

\bibitem{rei02}
P.~Reimann, Brownian motors: noisy transport far from equilibrium, Phys. Rep.
  361 (2002) 57.

\bibitem{asthan02}
R.~D. Astumian, P.~H{\"{a}}nggi, Brownian motors, Phys. Today 55 (2002) 33--39.

\bibitem{lined02}
in: H.~Linke (Ed.), Special issue on ratchets and brownian motion: basics,
  experiments and applications, Appl. Phys. A (55) v. 2, 2002.

\bibitem{reigrihan97}
P.~Reimann, M.~Grifoni, P.~H{\"{a}}nggi, Quantum ratchets, Phys. Rev. Lett. 79
  (1997) 10.

\bibitem{schvin02}
S.~Scheidl, V.~M. Vinokur, Quantum {B}rownian motion in ratchet potentials,
  Phys. Rev. B 65 (2002) 195305.

\bibitem{yuketal97}
S.~Yukawa, M.~Kikuchi, G.~Tatara, H.~Matsukawa, Quantum ratchets, J.~Phys.
  Society of Japan 66 (1997) 2953.

\bibitem{grietal02}
M.~Grifoni, M.~S. Ferreira, J.~Peguiron, J.~B. Majer, Quantum ratchets with few
  bands below the barrier, Phys. Rev. Lett. 89 (2002) 146801.

\bibitem{macetal04}
L.~Machura, M.~Kostur, P.~H{\"{a}}nggi, P.~Talkner, J.~Luczka, Consistent
  description of quantum brownian motors operating at strong friction, Phys.
  Rev. E 70 (2004) 031107.

\bibitem{peggri04}
J.~Peguiron, M.~Grifoni, Duality relation for quantum ratchets,
  cond-mat/0407759.

\bibitem{gar04}
J.~L. Garc{\'{\i}}a-Palacios, Solving quantum master equations in phase space
  by continued-fraction methods, Europhys. Lett. 65 (2004) 735.

\bibitem{garzue04}
J.~L. Garc{\'{\i}}a-Palacios, D.~Zueco, Caldeira-leggett quantum master
  equation in wigner phase space: continued-fraction solution and application
  to brownian motion in periodic potentials, J.~Phys. A 37 (2004) 10735.

\bibitem{weiss}
U.~Weiss, Quantum Dissipative Systems, World Scientific, Singapore, 1993.

\bibitem{calleg83pa}
A.~O. Caldeira, A.~J. Leggett, Path integral approach to quantum {B}rownian
  motion, Physica A 121 (1983) 587--616.

\bibitem{hiletal84}
M.~Hillery, R.~F. O'Connell, M.~O. Scully, E.~P. Wigner, Distribution functions
  in physics: {F}undamentals, Phys. Rep. 106 (1984) 121--167.

\bibitem{risken}
H.~Risken, The {F}okker--{P}lanck Equation, 2nd Edition, Springer, Berlin,
  1989.

\bibitem{borcosmar02}
M.~Borromeo, G.~Constantini, F.~Marchesoni, Deterministic ratchets: Route to
  diffusive transport, Phys. Rev. E 65 (2002) 041110.

\bibitem{shupol03}
A.~I. Shushin, E.~Pollak, Quantum and classical aspects of activated surface
  diffusion, J.~Chem. Phys. 119 (2003) 10941.

\bibitem{broozo04}
O.~Brodier, A.~M.~O. de~Almeida, Symplectic evolution of {W}igner functions in
  {M}arkovian open systems, Phys. Rev. E 69 (2004) 016204.

\bibitem{zur03RMP}
W.~H. Zurek, Decoherence, einselection, and the quantum origins of the
  classical, Rev. Mod. Phys. 75 (2003) 715.

\bibitem{sto03}
J.~T. Stockburger, Stochastic and numerical approaches to dissipative quantum
  dynamics: path integrals and beyond, Phys. Stat. Sol. 237 (2003) 146.

\end{thebibliography}

\end{document}